\documentclass[prd, twocolumn, nofootinbib, floatfix]{revtex4-1}
\pdfoutput=1

\usepackage{hyperref}
\usepackage{amsmath}
\usepackage{amssymb}
\usepackage{graphicx}
\usepackage{float} 
\usepackage{color}
\usepackage{multirow}

\usepackage[title]{appendix} 
\usepackage[dvipsnames]{xcolor}

\hypersetup{
    colorlinks=true,
    linkcolor=red,
    citecolor=blue,
}

\newcommand{\be}{\begin{equation}}
\newcommand{\ee}{\end{equation}}
\newcommand{\bea}{\begin{eqnarray}}
\newcommand{\eea}{\end{eqnarray}}
 
\newcommand{\dz}{\delta z}
\newcommand{\dzcore}{\delta z_{\rm core}}
\newcommand{\dzout}{\delta z_{\rm out}} 
\newcommand{\fout}{f_{\rm out}}
\newcommand{\ztrue}{z_{\rm true}}
\newcommand{\zphot}{z_{\rm phot}}

\definecolor{mediumpurple}{rgb}{0.58, 0.44, 0.86}

\begin{document}

\title{Cosmology Requirements on Supernova Photometric Redshift Systematics for\\ 
Rubin LSST and Roman Space Telescope} 

\author{Ayan Mitra} 
 \email{ayan.mitra@nu.edu.kz}
\affiliation{%
 School of Engineering and Digital Sciences, Nazarbayev University,\\ Nur-Sultan 010000, Kazakhstan} 
 
\author{Eric V.~Linder}
\affiliation{
Berkeley Center for Cosmological Physics \& Berkeley Lab, 
University of California, Berkeley, CA 94720, USA\\  
\\ 
Energetic Cosmos Laboratory, Nazarbayev University, Nur-Sultan 010000, Kazakhstan}

\date{\today} 

\begin{abstract}
Some million Type Ia supernovae (SN) will be discovered and 
monitored during upcoming wide area time domain surveys 
such as the Vera C.~Rubin Observatory Legacy Survey of Space and Time (LSST). For cosmological use, accurate redshifts 
are needed among other characteristics; however the vast 
majority of the SN will not have spectroscopic redshifts, 
even for their host galaxies, only photometric redshifts. 
We assess the redshift systematic control necessary for robust 
cosmology. Based on the photometric vs true redshift relation 
generated by machine learning applied to a simulation of 
500,000 galaxies as observed with LSST quality, 
we quantify requirements on systematics in the mean relation 
and in the outlier fraction and deviance so as not to bias 
dark energy cosmological inference. Certain 
redshift ranges are particularly sensitive, motivating 
spectroscopic followup of SN at $z\lesssim0.2$ and around 
$z\approx0.5$--0.6. Including Nancy Grace Roman Space Telescope 
near infrared bands in the simulation, we reanalyze the 
constraints, finding improvements at high redshift but little 
at the low redshifts where  systematics lead to strong 
cosmology bias. We identify a complete spectroscopic survey  
of SN host galaxies for $z\lesssim0.2$ as a highly favored 
element for robust SN cosmology. 
\end{abstract}

\maketitle

\section{\label{sec:level1}Introduction}

Type Ia supernovae (SN) are incisive probes of the cosmic expansion  
history, giving the tightest constraints on dark energy properties 
of any probe for a given distance precision. Wide field surveys can 
multiplex the observations, i.e.\ measure many SN at once in a given 
survey field, and time domain surveys that revisit fields on a 
cadence compatible with SN rise and fall times can monitor them 
throughout their lightcurve. Upcoming surveys such as ZTF \cite{ztf} 
and LSST \cite{lsst1,lsst2} are time domain surveys that are wide field (as well as 
deep, measuring SN over a significant range of redshifts) and with multiple 
wavelength bands, and thus will produce  
thousands to of order a million SN lightcurves. 

For best use as a cosmological probe, these sources need not only 
to be observed in multiple bands with good photometry for a 
significant part of their lightcurve (for accurate fitting and 
color corrections), as these surveys will provide, but also be 
classified as Type Ia supernovae to ensure a pure sample, 
and ideally subtyped as normal Type Ia. The last two characteristics 
are most robustly established through spectroscopy. The SN redshift 
must also be determined, e.g.\ through the redshift of the host 
galaxy, either through spectroscopy or photometry. 

This final step of the redshift determination for the  
distance-redshift relation to measure the cosmic expansion is what 
we focus on here, specifically the requirement on photometric 
redshift (`photo-z') accuracy. Statistical uncertainties in the redshift will 
propagate through to increasing the cosmological parameter 
uncertainties (see \cite{m0} for detailed calculations), 
but systematic errors will bias the cosmology. We extend the 
analysis of \cite{f1}, which employed analytic toy models for 
redshift systematic uncertainties, to use data simulated to 
reflect LSST observing characteristics, i.e.\ filters, exposure 
depths, etc. 

In Sec.~\ref{c1} we review the methodology for propagating systematic 
redshift errors into cosmological parameter bias. The simulated data and the 
derived photo-z vs true redshift mapping is discussed in Sec.~\ref{cc2}, 
with the results presented in Sec.~\ref{c3}, along with the requirements 
necessary to avoid significant bias. 
We explore the effects of external near infrared data (NIR) from  
the Roman Telescope in Sec.~\ref{sec:roman}, and 
conclude in Sec.~\ref{c4}.

\section{Method}\label{c1} 

To propagate an observational systematic to bias in cosmology 
inference, a well used and convenient technique is the Fisher  
information bias formalism \cite{9805012,0604280}. Specifically 
we follow the approach in \cite{f1}, with a parameter set including 
the supernova absolute magnitude parameter $\mathcal{M}$, the 
present matter density $\Omega_m$ as a fraction of the critical density, 
and the dark energy equation of state parameters $w_0$ and $w_a$ 
giving its present value and a measure of its time variation. 
We include the effects of a misestimated redshift $z$ on both the 
distances and the lightcurve width-luminosity relation (but not 
extinction, which is expected to be a small effect). 

The final relation, as in \cite{f1}, for the apparent 
magnitude offset $\Delta m$ is 
\begin{equation}\label{delta_m}
    \Delta m = \frac{5}{\ln 10}\ln{\left[ \frac{D_L(z+\delta z)}{D_L(z)} \right]} +1.4\, \frac{\delta z}{1+z}\ , 
\end{equation}
where $\delta z$ is the redshift systematic at redshift $z$ 
and $D_L$ is the luminosity distance. This misestimation 
$\Delta m$ then biases the cosmology parameters. 
Imposing constraints on the degree of cosmology bias in turn  
propagates back to requirements on redshift systematics, as a  
function of type, degree, and redshift at which they occur. 
We will present the confidence contours in the dark energy  
$w_0$--$w_a$ plane (marginalized over the other parameters), 
and require that the bias shift the cosmology by less 
than $1\sigma$, i.e.\ staying within the 68.3\% joint 
confidence level contour: $\Delta\chi^2<2.3$. 

In \cite{f1} we focused on toy models for additive and 
multiplicative systematics, and outliers. Here we use 
the photometric vs true redshift mapping derived from 
simulations, as described in the next section.

\section{Data}\label{cc2} 

To obtain the redshift of a SN, the most common method is to measure 
the host galaxy redshift. This either exists in previous catalogs 
or can be determined by the survey itself, possibly after the SN 
has faded. Less common is getting an estimate of the redshift from 
the SN colors (flux differences between wavelength bands) 
\cite{kessler}; it is also possible that this could be useful in 
avoiding catastrophic outliers in the galaxy photometric redshift 
\cite{kessler2,dai,zbeams}. 
However this has not been fully tested for side effects and confirmed, 
and here we consider only host galaxy redshifts.

\subsection{Photo-z Catalog} \label{sec:catalog} 

The leading current simulations of LSST galaxy photometric 
redshift distributions use  
color matched nearest neighbors (CMNN) estimators 
\cite{m1,m2}. This incorporates the expected 
photometric data quality and survey characteristics. 
The photo-z estimator is based on the location of a 
test galaxy in color space, identifying it to the nearest 
color matched galaxy from the spectroscopic training data set. 
Minimization on the distance metric is done via $\chi^2$ to 
estimate the photo-z of the test galaxy. The $\chi^2$ 
(Mahalanobis) distance $D$ is computed as 
\be
D = \sum_1^{N}\frac{(c_{\rm train}-c_{\rm test})^2}{(\delta c)^2}\,, 
\ee
where $c$ is the color, $N$ is the total number of colors, 
and $\delta c$ quantifies the measurement error in color. 
A test galaxy needs to be detected with at least $N=3$ colors 
to be assigned a photo-z. 

The training set is analogous to 
the spectroscopic galaxy sample 
while the test set is composed of galaxies with simulated 
colors, for which the photo-z are estimated. For training the 
CMNN estimator we have simulated a larger data set than in  
\cite{m2}, of  500,000 galaxies 
and for testing a sample size of 90,000.  As both are 
simulated, the true redshifts are known and the distribution  
of the derived photo-z relative to the true redshift can be 
mapped. 

This estimator is designed such that the accuracy and the precision of the photo-z are directly related to the precision of the survey's photometry. 
The training and testing sets were drawn entirely from the simulated photometry data catalog, with both having the same distribution of redshift and magnitude (flux). 
The simulated galaxy catalog is based on the Millennium simulation \cite{mil} and uses realistic photometric characteristics. 
Details on the galaxy catalog construction are described in \cite{gonz,mers}. 

The CMNN photo-z estimator was designed to model the optical ($ugrizy$) and NIR ($YJHK$) properties of galaxies. 
For the  NIR, note that from the figures in \cite{m1} 
the NIR bands from the Euclid 
satellite \cite{euclid} do not significantly impact the 
photo-z systematic uncertainties at redshifts 
$z\lesssim1$ where we observe SN with LSST 
(note this refers to the much tighter  photo-z requirements for supernovae; Euclid will be quite valuable for weak lensing photo-z constraints), though those from the 
Nancy Grace Roman Space Telescope 
\cite{ngrst} could. 
For the purposes of this work we use only the $10$ year LSST projections, with the optical filters' $5\sigma$ detection limits as  tabulated in 
Table~\ref{T1}, as the main input. In Section~\ref{sec:roman} we 
extend this to include Roman $YJH$ bands.

 \begin{table}
        \begin{center}
            \begin{tabular}{ c c c c c c c c c c}
                $u$ & $g$ & $r$ & $i$ & $z$ & $y$ 
                \\[.3em]
                \hline
                26.1 & 27.4 & 27.5 & 26.8 & 26.1 & 24.9 
                \\ 
            \end{tabular}
        \end{center}
        \caption{Summary of the $5\sigma$ limiting magnitudes for each filter for a LSST $10$ year forecast, based on the LSST simulation software package \cite{connolly}.  
        }
        \label{T1}
    \end{table}

\subsection{Photo-z Systematics} 
\label{photozs} 

Given the catalog of photo-z's and true redshifts, one can 
carry out various statistical analyses to assess robustness 
of the distribution. 
Ref.~\cite{m2} presented different statistical measures of the photo-z quality based on the results from the CMNN estimators on the mock galaxy catalog. In this analysis, we will follow 
these definitions, with alterations as described in 
Sec.~\ref{c3}. 
From these we will derive the quantity needed for the cosmology 
requirements on systematics, $\delta z(z)$ in  Eq.~\eqref{delta_m}. 

To model the LSST-like uncertainty, Gaussian random photometric 
scatter was added to the simulated observed apparent magnitudes 
from the true catalog with a standard deviation equal to the 
predicted magnitude error for each galaxy. The magnitude error was modeled as for LSST, based on the description in \cite{ivezic}.
For computing the cosmology bias we use two types of  
photo-z systematic offsets for $\delta z$, the bias  
in the core of the photo-z distribution (referred to 
in \cite{m2} as ``robust bias'') and in the outlier 
distribution.

\subsubsection{Robust Bias} 

If the mean photo-z relation not only scatters about 
the true redshift but is biased from it, generally by  
different amounts at different redshifts, this will 
lead to a cosmology bias. This core offset, or robust 
bias, is defined as the mean bias in the inter quartile range 
(IQR; defined as including 50\% of the galaxies) 
of the galaxy photo-z error \cite{m2}. For a photo-z error 
defined as 
\begin{equation}
\Delta z_{(1+z)}=\left(\frac{\ztrue-\zphot}{1+\zphot}\right) \ , 
\end{equation}
where $\ztrue$ is the true or spectroscopic catalog redshift and $\zphot$ is the estimated photometric redshift, 
the robust bias is taken to be the mean over the inter quartile range 
\begin{equation} 
{\rm Robust\ Bias}=\overline{\Delta z_{(1+z),{\rm IQR}}} \ . 
\end{equation}  
We refer to this as $\delta z_{\rm core}$.

\subsubsection{Outlier Bias} 

Photo-z errors sometimes have large departures from 
the true values, lying outside the core. There are two 
types of these: outliers and catastrophic outliers. 
Here we need to know not only the degree of offset but 
the fraction of  photo-z's that are outliers. 

Catastrophic outliers are defined as 
$|\ztrue-\zphot|>1.5$ and it is highly unlikely 
that a SN with such a mistaken redshift would go  
unrecognized and be placed on a Hubble diagram (its 
peak magnitude, lightcurve width, etc.\ would lie well 
off expectation for a reasonable variation of cosmology). 
Therefore we do not consider catastrophic outliers. 

We classify outliers as photo-z's that lie outside the 
core, defined as $\Delta z_{(1+z)}>3\sigma(z)$ where 
$\sigma(z)$ is the standard deviation for the IQR 
galaxies. Furthermore we require $\Delta z_{(1+z)}>0.06$.

\subsubsection{Sample Selection}\label{ss} 

To select our data set for the analysis, we modified and 
reran the CMNN 
simulation with two alterations relative to \cite{m2}: 
\begin{itemize}
\item $0<\ztrue\le1.2$, as the LSST supernova survey is expected to cover this range, i.e.\ SN  at higher true 
redshifts would likely be too faint for the survey 
magnitude limits. This restricts the redshift range on 
the high end  relative to \cite{m2}, but expands it on 
the low end since \cite{m2} did not identify outliers for 
$z\le 0.2$. 
\item $|\delta z|<1.5$, i.e.\ we distinguish between 
outliers and catastrophic outliers. 
\end{itemize}

\begin{figure}
\includegraphics[width=\columnwidth]{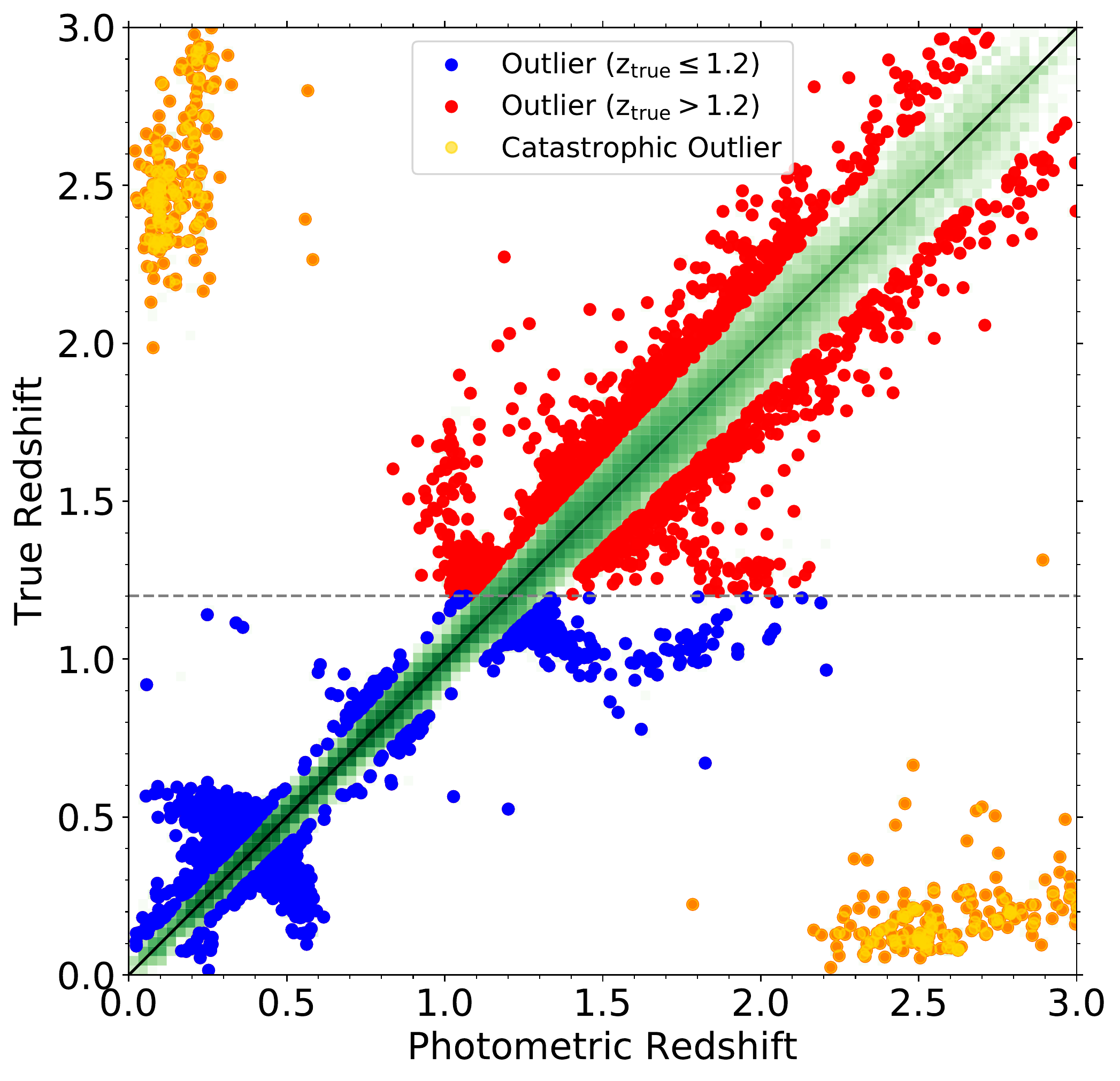}
\caption{
Photo-z's vs true redshifts resulting from the CMNN 
galaxy simulation \cite{m2}, with modified selection 
criteria. Galaxies are classified as lying in the core 
of the distribution (green), outliers (blue if 
$\ztrue\le1.2$, red if $\ztrue>1.2$), or 
catastrophic outliers (gold). The grey horizontal dashed  
line shows the $\ztrue\le1.2$ cut and the black diagonal 
line shows a perfect survey, with $\zphot=\ztrue$. 
Any deviation in the mean from the diagonal gives a bias.  
} 
\label{N}
\end{figure}

Figure~\ref{N} shows the result of the updated simulation 
analysis. Core points are in green, outliers are in blue 
if they fall within the $\ztrue$ cut and red if they are 
at higher true redshift, while gold points are catastrophic 
outliers and not included in the cosmology analysis. 
Green core points out to $\ztrue\le1.2$ determine 
$\delta z_{\rm core}(\ztrue)$ while blue points determine 
$\delta z_{\rm out}(\ztrue)$ and the fraction of outliers  
$f_{\rm  out}(\ztrue)$, equaling the ratio of the 
number of blue points to blue+green points at that 
redshift. 
We use a binning on the statistics of 
$\Delta z=0.05$, much finer than the original 
$\Delta z=0.3$. 
To train the CMNN estimator, we simulated 500,000 galaxies for training and 90,000 galaxies for testing (out to $\ztrue=3$).

\section{Analysis}\label{c3} 

We propagate the quantities $\dzcore$, $\dzout$, and $\fout$ 
derived from analysis of Fig.~\ref{N} into the cosmology 
analysis, i.e.\ Eq.~\eqref{delta_m} and then the Fisher  
information analysis. 
The quantity $\fout$ enters there 
since only a fraction $\fout$ of the SN have the ensuing 
outlier bias $\Delta m$. Thus, both $\dzout$ and $\fout$ 
are important: if the redshift offset $\dzout$ is high, but 
happens only rarely (low $\fout$), this will give a small 
cosmology bias, as will a large outlier fraction $\fout$ 
but with only a small offset $\dzout$. 

For the cosmology calculation we follow the analysis 
described in \cite{f1}, with SN distributed over the 
range $z=[0-1.2]$. We evaluate $\delta z$ in bins of 
width 
$\Delta z=0.05$ (see 
discussion below). We treat the core bias and outlier 
bias separately, for clarity. 

\begin{figure}
    \includegraphics[width=\columnwidth]{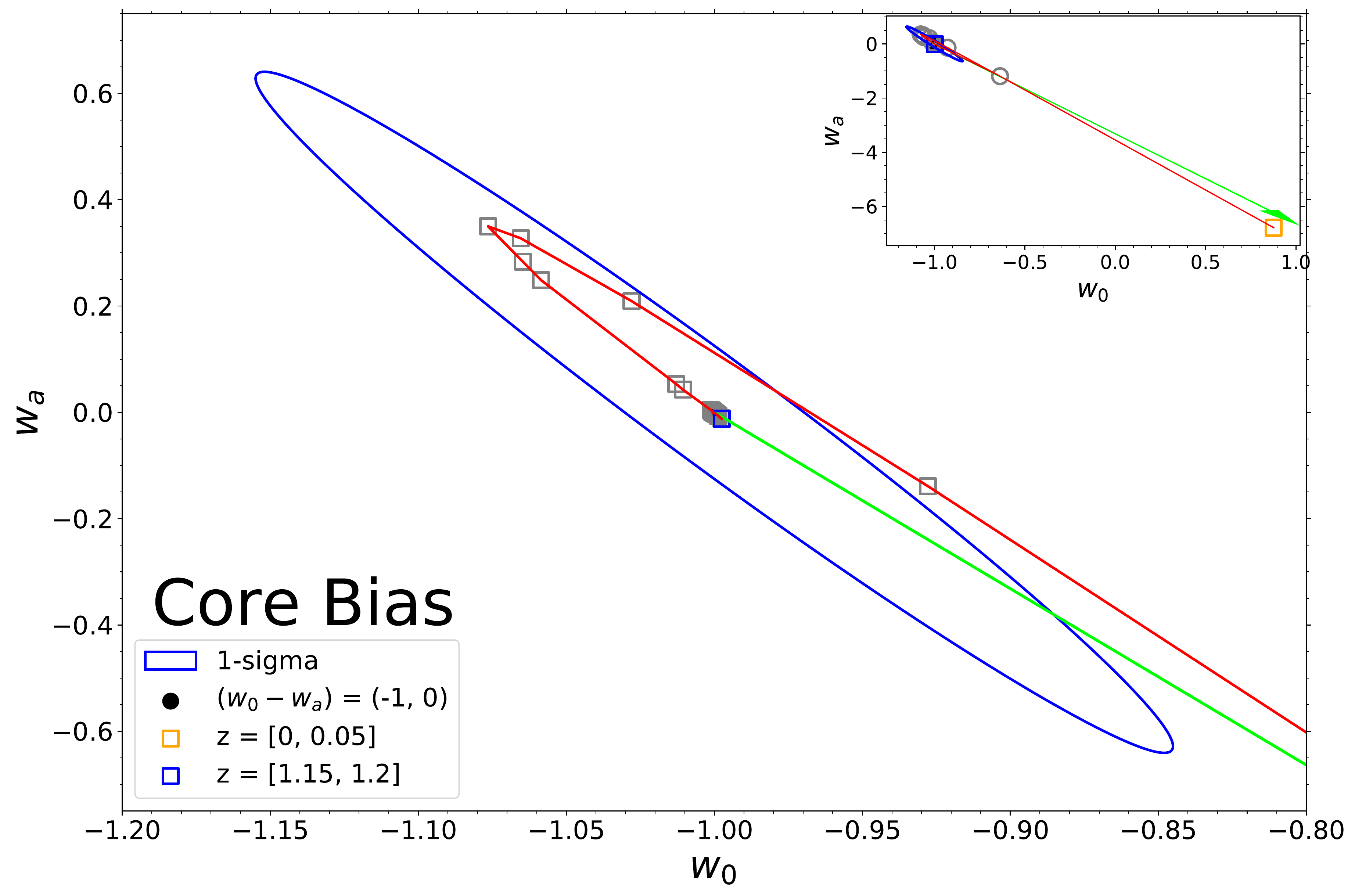}%
    \caption{ 
Cosmology bias in the dark energy $w_0$--$w_a$ plane 
coming from core bias redshift systematics 
is shown by the red curve tracing the effect from 
one redshift bin systematic $\dzcore(z)$ at a time, marked 
by squares from the lowest (orange) to highest  
redshift (blue). The green arrow gives the vector sum 
over all redshifts, with the inset showing the full 
extent of the bias. For reference, the statistical 
uncertainty is shown as the $1\sigma$ (68.3\%) joint 
confidence contour (blue ellipse). 
} 
    \label{N2}
\end{figure}

Figure~\ref{N2} shows the bias in cosmology as a result 
of the core bias redshift systematic. The input 
cosmology has $(w_0,w_a)=(-1,0)$ and the ellipse shows 
the $1\sigma$ (68.3\%) joint confidence contour, 
marginalized over the other parameters (matter density 
and SN absolute magnitude ${\mathcal M}$), for a rough 
approximation of the LSST SN sample plus a Planck CMB 
prior on the distance to last scattering. 

Applying the $\dzcore(z)$ systematic to the redshift bins 
one by one, the corresponding shift in the cosmology 
is shown by the individual square boxes, beginning 
with the orange box ($z=[0,0.05]$), and ending with the 
blue box ($z=[1.15,1.2]$). The red line connecting the 
boxes traces the locus of the cosmology bias with 
increasing bin redshift. The green arrow shows the net 
effect (vector sum) for the systematic present on all 
the redshift bins. Note the direction of bias from 
some redshifts is such that it can diminish bias from 
another redshift. Cosmology is most sensitive to the 
low redshift systematics, with systematics in the three 
lowest bins biasing cosmology outside the $1\sigma$ 
contour. This can be a hopeful sign in that these 
can be the most easily addressed with supplementary 
observations.

Next we turn to the outlier bias. For this we find the  
fine redshift bins of width $\Delta z = 0.05$ to  be important to treat properly the sharp 
outlier features, especially at low redshift. 
Quantitatively, the cosmology parameter bias is 
misestimated by using bins of $\Delta z = 0.1$ rather  
than 0.05 by $\sim0.03$ in $w_0$ and 0.13 in $w_a$ 
only at $z<0.2$; above this the maximum errors are 
$\lesssim0.015$ and 0.06 respectively, and generally 
much less. 
Recall that for redshift systematics due to outliers we care about 
both the offset and the fraction of galaxies exhibiting the 
systematics. Except at low redshift (where $dm/dz$ is steep), the 
cosmology bias is basically proportional to their product, 
$\fout\,\delta z$. 

\begin{figure}
    \includegraphics[width=\columnwidth]{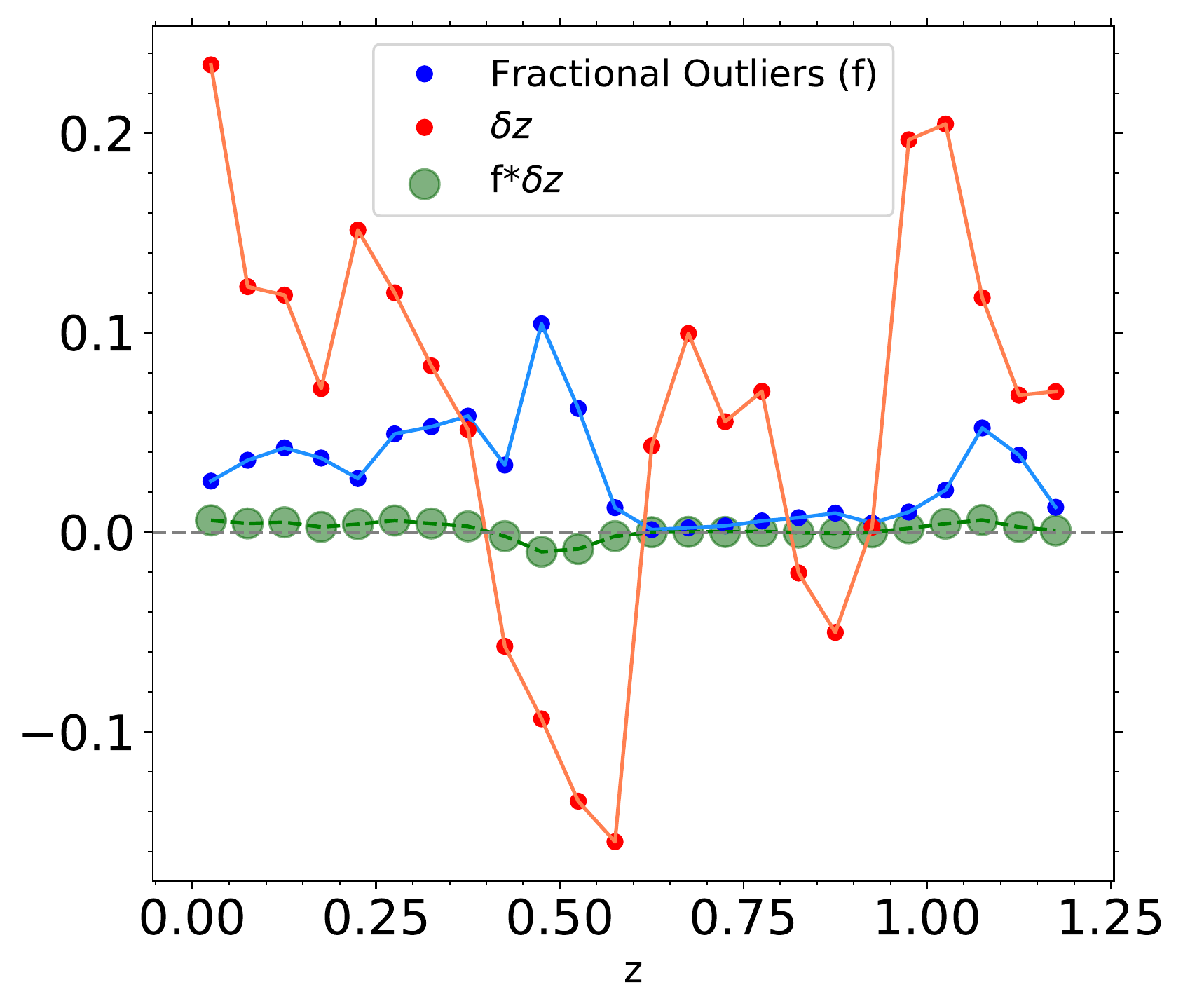}
    \caption{Simulation output for outlier redshift systematics 
    magnitude $\dzout(z)$, fraction of outliers $\fout(z)$, and their 
    product. Sharp features in the outlier systematics require 
    fine binning in redshift. 
}
    \label{N3+}
\end{figure}

Figure~\ref{N3+} shows the simulation results for $\dz(z)$, 
$\fout(z)$, and their product. Although the product looks quite 
small, one must propagate it to the cosmology bias to determine  
its impact. Even a small systematic can have a significant effect 
on a high precision survey such as LSST.

\begin{figure}
    \includegraphics[width=\columnwidth]{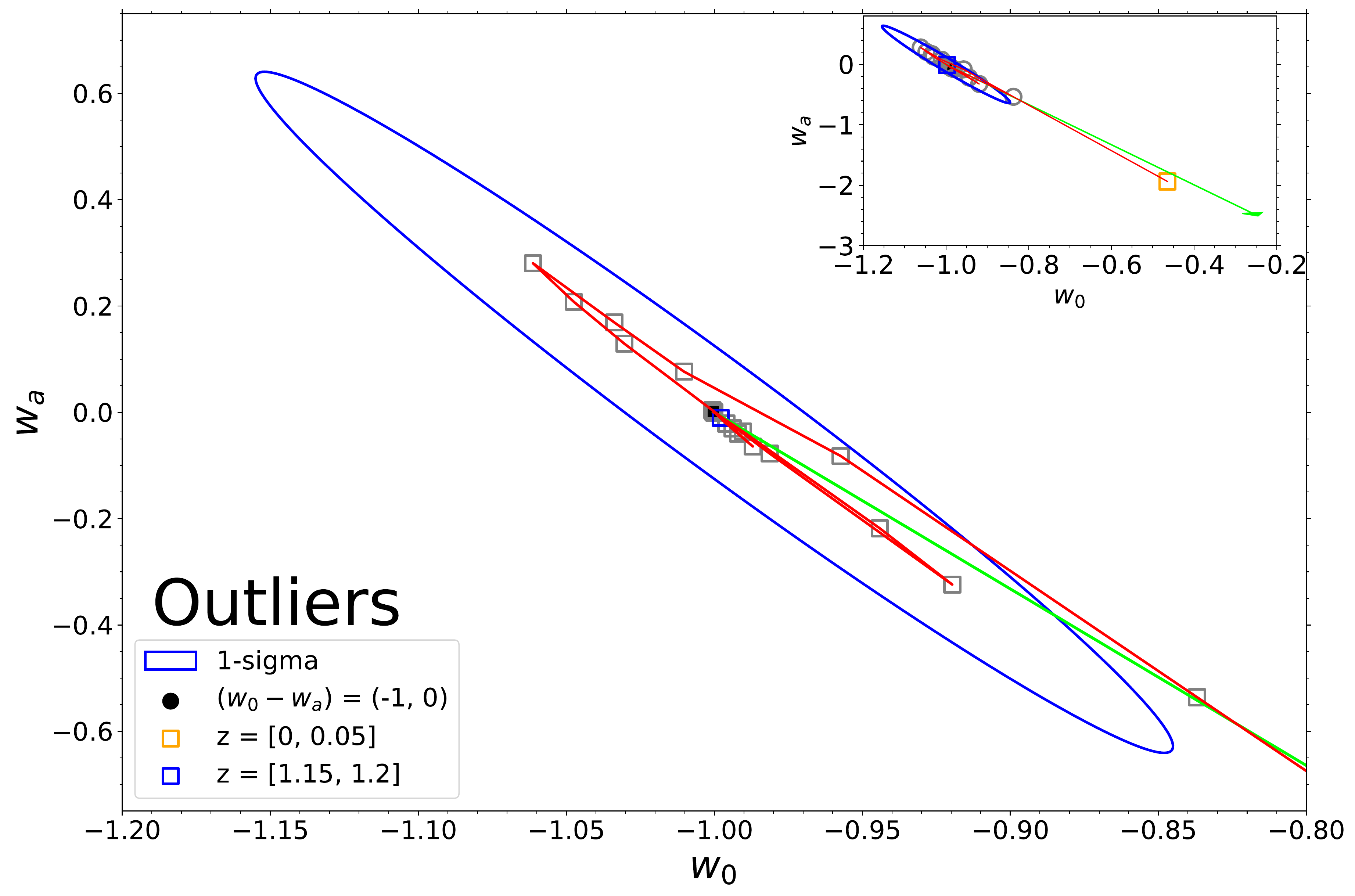}%
    \caption{As Figure~\ref{N2}, but for outlier bias. Again, 
    low redshift bins $z\lesssim0.2$ are particularly affected. 
    }
    \label{N3}
\end{figure}

Figure~\ref{N3} shows the resulting cosmology bias from the outlier 
systematics. While the effect on the first redshift bin is less than 
from the core bias, systematics from the remainder of the redshift 
range have comparable, significant effects. Of course both types of 
systematics will be present, and their induced bias goes in the same 
direction for distorting the cosmology inference.

Since the cosmology bias from the redshift core systematics 
and the outlier systematics, and their sum, are well beyond 
the desired $1\sigma$ statistical confidence contour, it must be reduced for 
useful cosmology estimation from LSST photometric supernovae. 
We present three possible strategies for amelioration: 1) modeling, 
2) select spectroscopy, 3) external imaging data. 

\begin{figure}
    \includegraphics[width=\columnwidth]{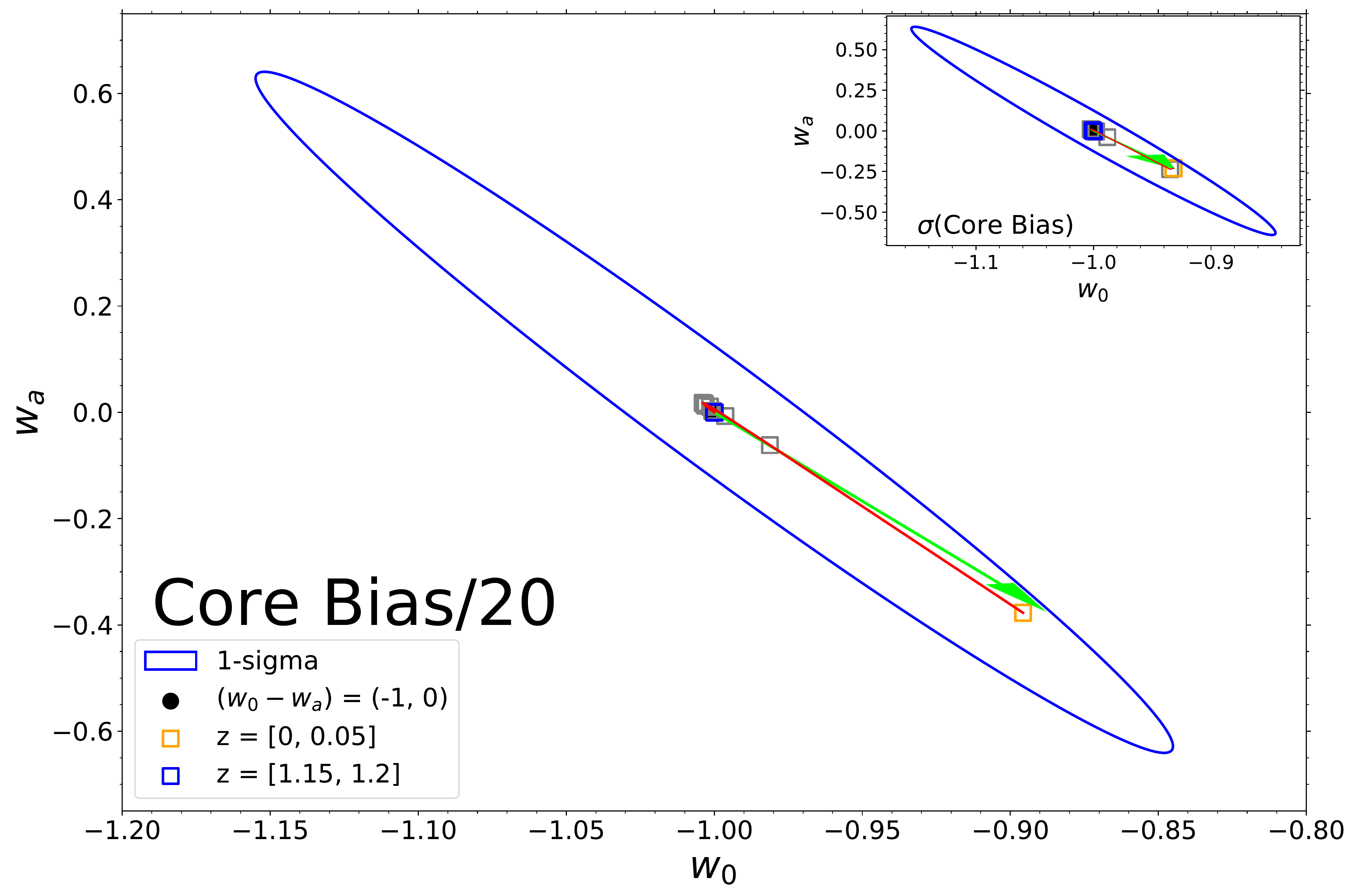}
    \caption{
    As Figure~\ref{N2}, but with the core bias now scaled 
    down by a factor 20, mimicking modeling of the systematic to leave 
    a small residual. At this level of control, the net bias is contained 
    within the $1\sigma$ statistical confidence contour. The inset 
    shows the result if the systematics were modeled down to the bootstrap sampling  error from 1000 simulations. 
}
    \label{N4}
\end{figure}

The redshift systematics found by the simulations is a raw systematic, 
without remediation. One could attempt to model the systematics and 
correct for them, up to the fidelity of the modeling process, leaving 
a smaller, residual systematic. 
Figure~\ref{N4} shows that if the residual core bias is scaled down 
by a factor of 20, i.e.\ leaving only 5\% of the simulation systematic, then the cosmology bias lies within the $1\sigma$ confidence contour. 
(In fact, one should not take the factor 20 too precisely: 
the Fisher bias formalism is valid for small changes in 
the observable, i.e.\ $\Delta m$, so the effect of large 
changes at the lowest redshifts is possibly exaggerated.) 
The inset shows the case for the 
systematics reduced all the way to 
the bootstrap sampling error\footnote{The 
bootstrap error in the core bias is derived by randomly drawing galaxy subsets with replacement and recalculating the statistics 1000 times, and then using the standard deviation of all recalculations as the error.} 
\cite{m2}. A similar process could potentially be applied to the 
outlier bias.

The second strategy involves targeting the systematics in particular 
redshift ranges. We have seen that the low redshift systematics 
produce the highest cosmology bias. This is expected as the SN apparent magnitude $m$ on the Hubble diagram starts off steeply varying with 
redshift, roughly $\sim 1/z$, and then flattens at higher redshift. 
So a small $\dz$ at low redshift has a significant effect. 
This is fortunate in that the low redshift region is most amenable 
to use of spectroscopy to determine the SN redshift (e.g.\ through 
targeting its host galaxy). 

\begin{figure}
    \includegraphics[width=\columnwidth]{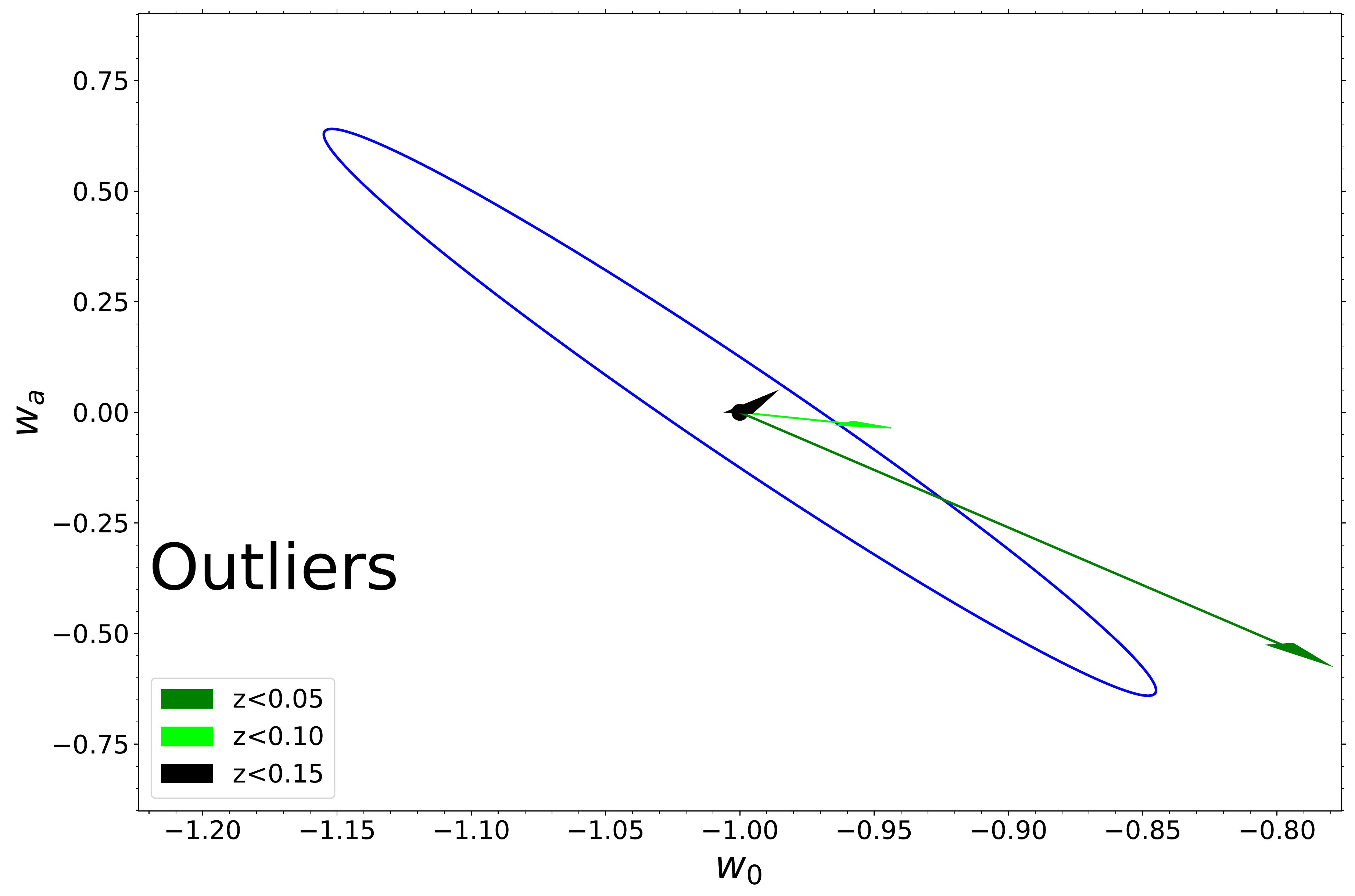}%
    \caption{As Figure~\ref{N3}, but with spectroscopic redshifts, 
    and hence $\fout=0$, for $z<z_\star$, for three different values 
    of $z_\star$. 
   }
    \label{N5}
\end{figure}

Figure~\ref{N5} shows the application of this approach to the 
case of outlier systematics. The three arrows correspond to the total 
cosmology bias that ensues from SN over the full range $z=[0,1.2]$ if 
those (and only those) SN at $z<z_\star$ get spectroscopic 
redshifts, and so there are no outliers there: $\fout(z<z_\star)=0$. 
Spectroscopic redshifts for $z\lesssim0.2$ SN bring the bias under 
control. One possibility for carrying this out is through the 
secondary target program of the DESI Bright Galaxy Survey \cite{bgs}, 
or other multifiber spectrographs  \cite{mos,elt}. Such a low redshift 
spectroscopic SN sample is quite interesting scientifically as it also has power as a cosmic probe of gravity  
through peculiar velocities \cite{pecvel,peculiar}.

We consider the third strategy of utilization of external imaging data 
to improve redshift systematics in the next section.

\section{LSST + Roman Space Telescope} \label{sec:roman} 

During the LSST, supplementary imaging data that can help constrain the 
photometric redshift estimation will be available, notably from the Euclid 
satellite and the Nancy Grace Roman Space Telescope, both with near  
infrared wavelength measurements. As discussed in Section~\ref{sec:catalog}, 
Euclid's leverage comes at redshifts higher than those whose photometric 
systematics most significantly impact the supernova cosmology. Roman, 
however, extends to lower redshifts and we consider the benefit from 
adding YJH\footnote{The K 
filter, really F184, is  useful at 
much higher redshift than we consider here. 
See Fig.~10 and Sec.~4 in \cite{m2} for 
discussion.} 
imaging data with exposure depths comparable to those from 
Roman to the photometric redshift determination  (26.7, 26.9, 26.0 mag 
respectively for $5\sigma$ limiting depths). We follow  
\cite{m2} for the joint analysis photo-z catalogs. While 
Roman will of course have its own highly incisive sample of spectroscopic 
SN, here we consider only its effect on LSST photo-z. 

We carry out the joint analysis in two distinct ways. First, we consider 
the effect of the added Roman information on the outlier systematics 
analysis of Section~\ref{c3}, comparing LSST alone with LSST+Roman. 
However, since the addition of Roman data can decrease the photo-z 
standard deviation, the outlier fraction can actually increase since 
it is defined in terms of galaxy photo-z's deviating from $\ztrue$ 
by $3\sigma$. (Also, LSST catastrophic outliers can become 
LSST+Roman regular outliers.) Therefore we also discuss the full 
systematics -- from outlier galaxies (outside 
$3\sigma$) and inlier galaxies (within $3\sigma$) -- in a second cosmology analysis.

\subsection{LSST+Roman Outlier Analysis} \label{sec:lrout} 

The color matching nearest neighbors (CMNN) 
algorithm of \cite{m2} works somewhat differently 
when combining external data with LSST, so the 
interested reader should consult that paper for a  
full discussion. The main point to note here is that 
because including new filters not only adds information 
but also degrees of freedom, i.e.\ fit parameters, 
in the algorithm, if the additional NIR 
filter does not 
carry clear photo-z information (as can occur at, 
say, low redshift) then the combination can actually 
give worse results than LSST alone. Future work could consider how to 
treat this, either by cutting or tapering multiband 
information in such cases, or adjusting the CMNN  
algorithm. 

\begin{figure}
    \includegraphics[width=\columnwidth]{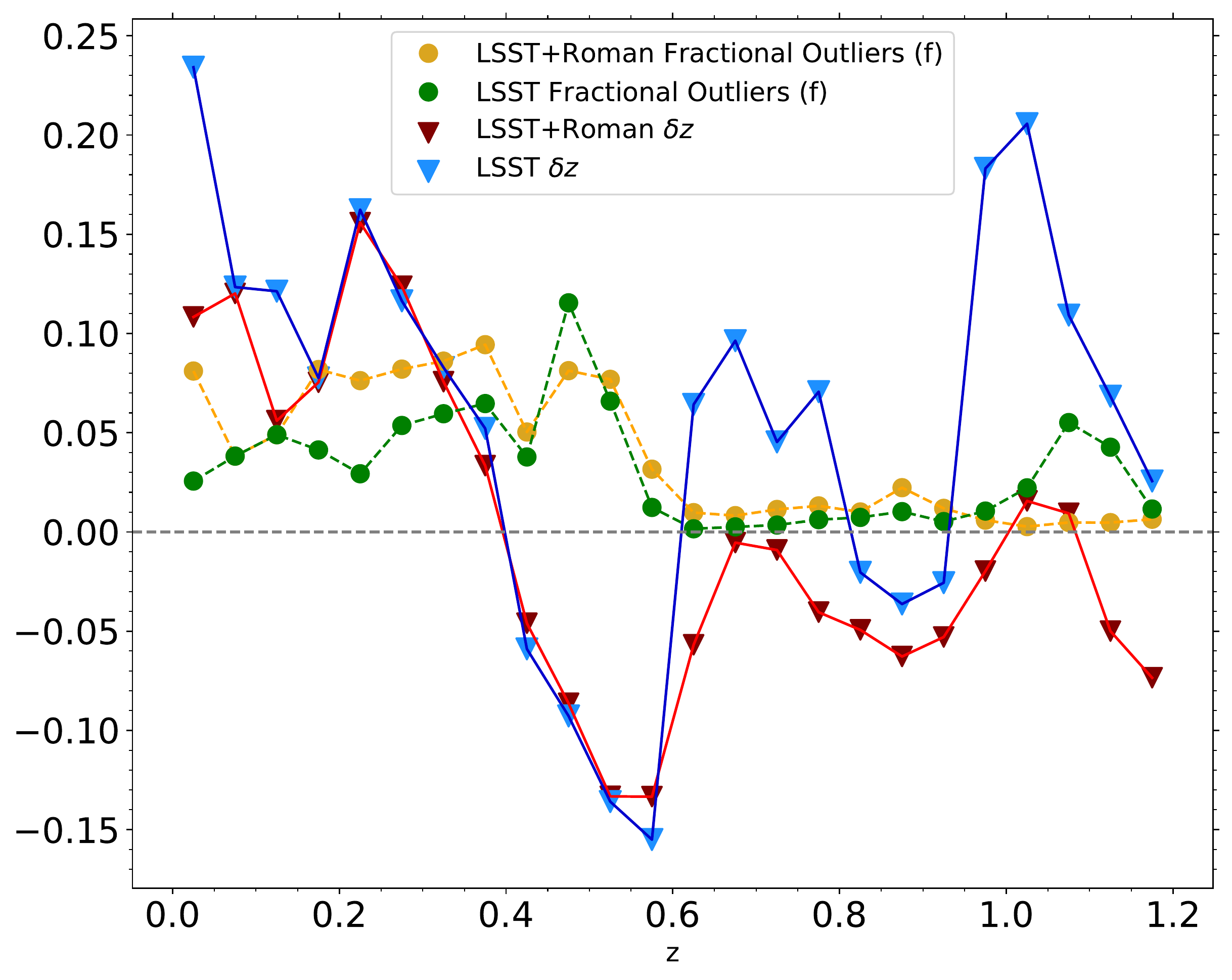}%
    \caption{Comparison between LSST+Roman and LSST   simulation output for outlier redshift systematics 
    magnitude $\dzout(z)$ and the corresponding  fraction of outliers $\fout(z)$. 
}

    \label{Va}
\end{figure}

Figure~\ref{Va} compares outlier systematics in terms 
of  $\dzout$ and $\fout$ for LSST and LSST+Roman. 
We see that Roman does help to tame the excess 
systematic in the lowest redshift bin, although it 
actually has a higher fraction of outliers 
there. Over the range $z=[0.2,0.6]$ there is little 
impact on $\dzout$ from the NIR bands, but Roman 
data improves the photo-z significantly for 
$z\gtrsim0.6$. For $z\gtrsim1$, the outlier 
fraction when including Roman is strongly reduced, 
so what deviations $\dzout$ do exist only affect a 
small number of galaxies.

\begin{figure}
    \includegraphics[width=\columnwidth]{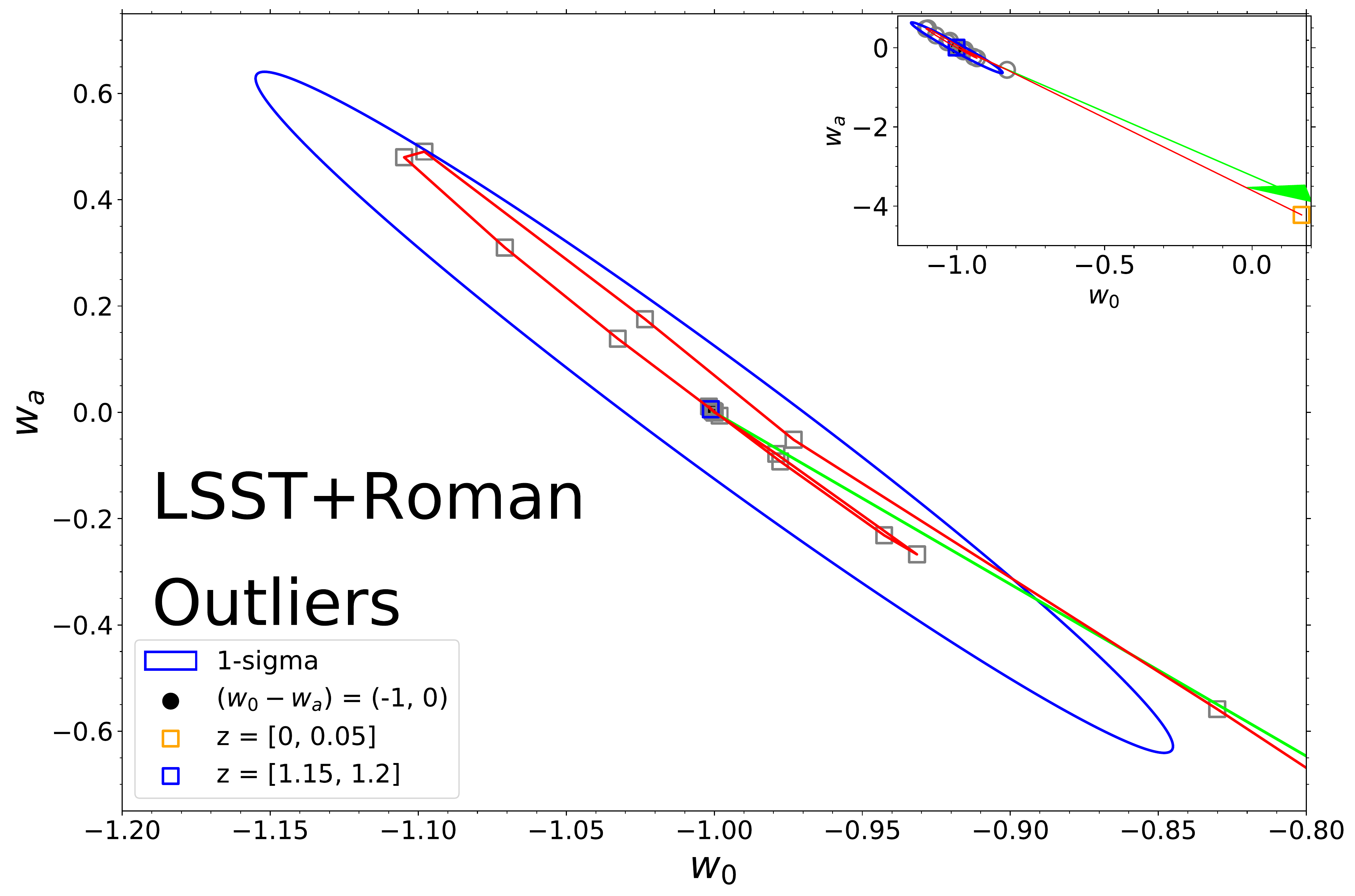}%
    \caption{As Figure~\ref{N3}, but cosmology bias from the outlier systematics in the LSST+Roman case. 
    }
    \label{Vc}
\end{figure}

The outlier systematics at $z\lesssim0.2$ and around 
$z\approx0.5$ still remain, so the cosmology bias issue  
is not solved. Figure~\ref{Vc} shows the cosmological effects 
of the photo-z systematic. By comparing to Figure~\ref{N3} we 
see qualitatively similar behavior, and the quantitative  
aspects are also not very different. The improvement in photo-z at 
high redshifts from Roman does not result in a large impact 
because high redshifts give a fairly modest contribution to 
the cosmology bias, which is much more sensitive to low 
redshifts. Thus, the need for low redshift spectroscopy to 
remove the photo-z outlier systematic remains strong.

\begin{figure}
    \includegraphics[width=\columnwidth]{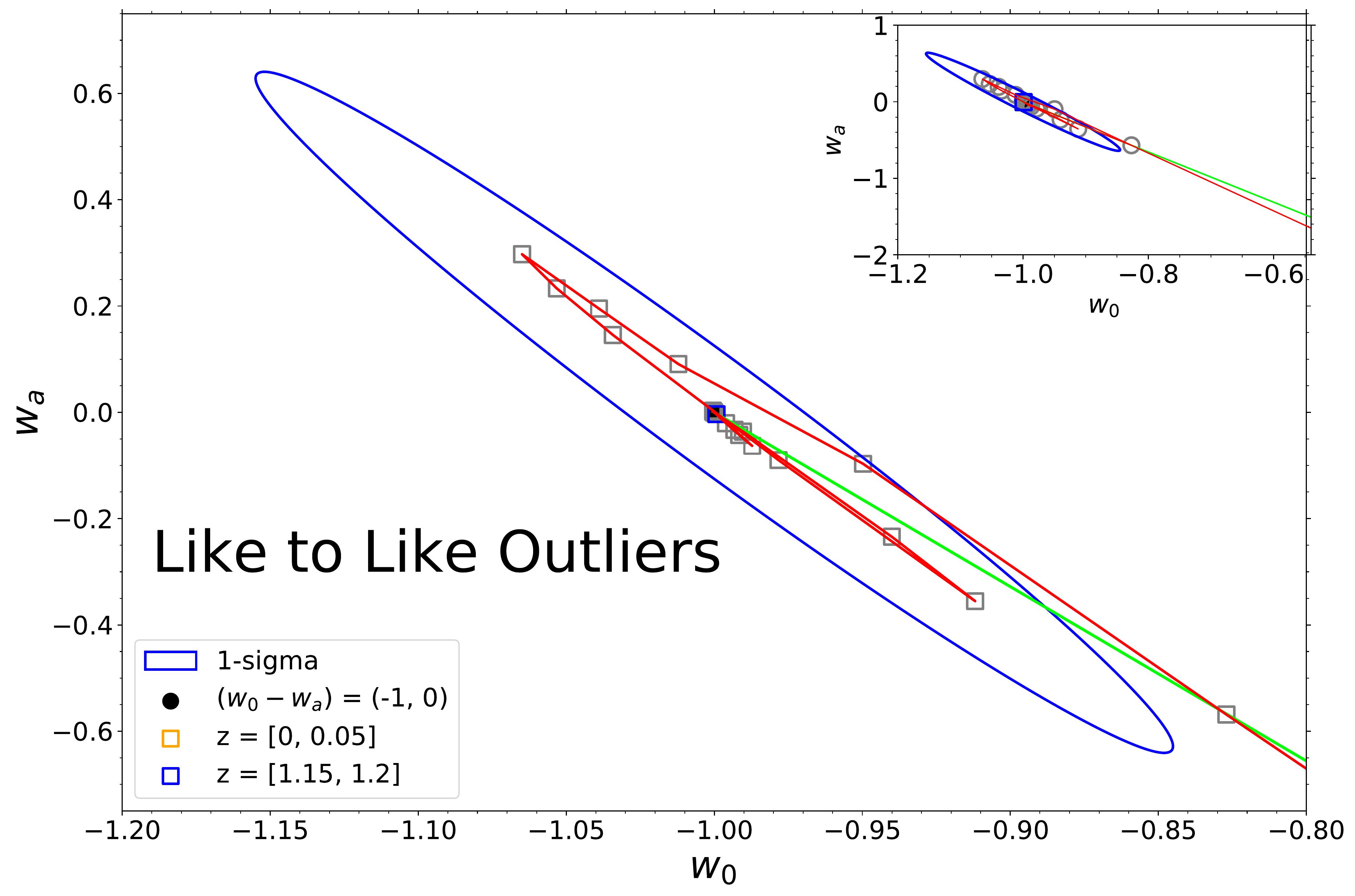}
\caption{As Figure~\ref{Vc}, but for a like to like treatment 
where the LSST+Roman outliers are restricted to galaxies that 
were also LSST alone outliers. Cosmology bias is reduced from 
LSST alone (Figure~\ref{N3}) and from LSST+Roman general outliers 
(Figure~\ref{Vc}). 
}
    \label{W1}
\end{figure}

As mentioned, adding Roman NIR data to LSST changes which galaxies 
are considered outliers, and so the ameliorating effects of the 
extra data are somewhat obscured. We can consider only those 
galaxies classified as outliers using just LSST, and then add 
Roman data to those alone and examine the properties of those that 
remain outliers: this is a ``like to like'' comparison. 
Figure~\ref{W1} shows the cosmology bias for this case, and we 
see that indeed significant improvement is evident.

\subsection{LSST+Roman All Galaxy Analysis} \label{sec:lrall} 

Adding NIR not only affects the photo-z outliers, but changes 
the size of the core and hence the dividing line between outliers 
and inliers. To take into account all the effects on the photo-z systematics 
of adding NIR data, we consider here all the galaxies 
together, outliers and inliers, and compute the total 
cosmology bias. 

\begin{figure}
    \includegraphics[width=\columnwidth]{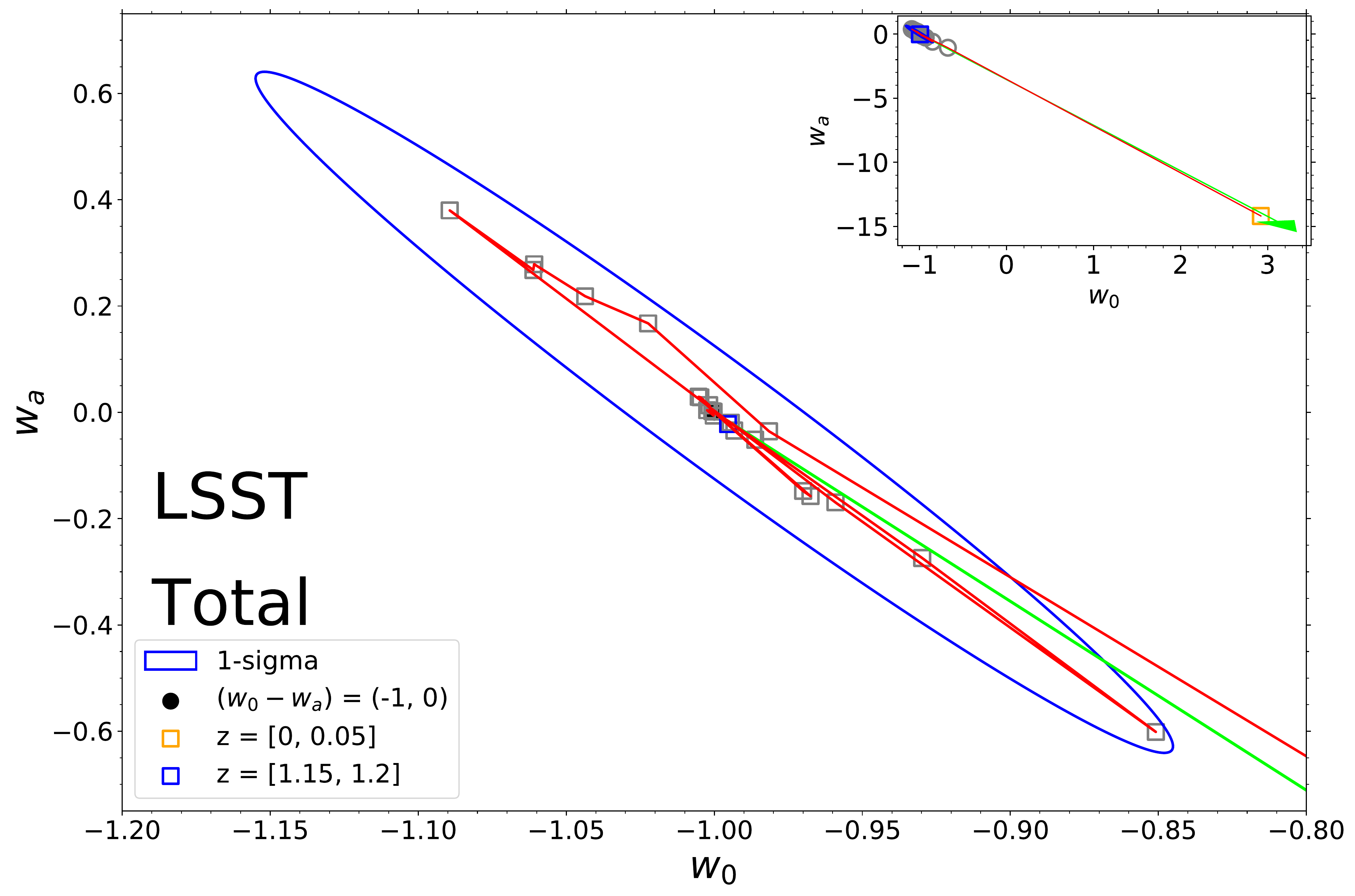}%
    \caption{As Figure~\ref{N2}, but 
    cosmology bias from the total outlier and inlier 
photo-z systematics for LSST. 
    }
    \label{Vd}
\end{figure}

\begin{figure}
    \includegraphics[width=\columnwidth]{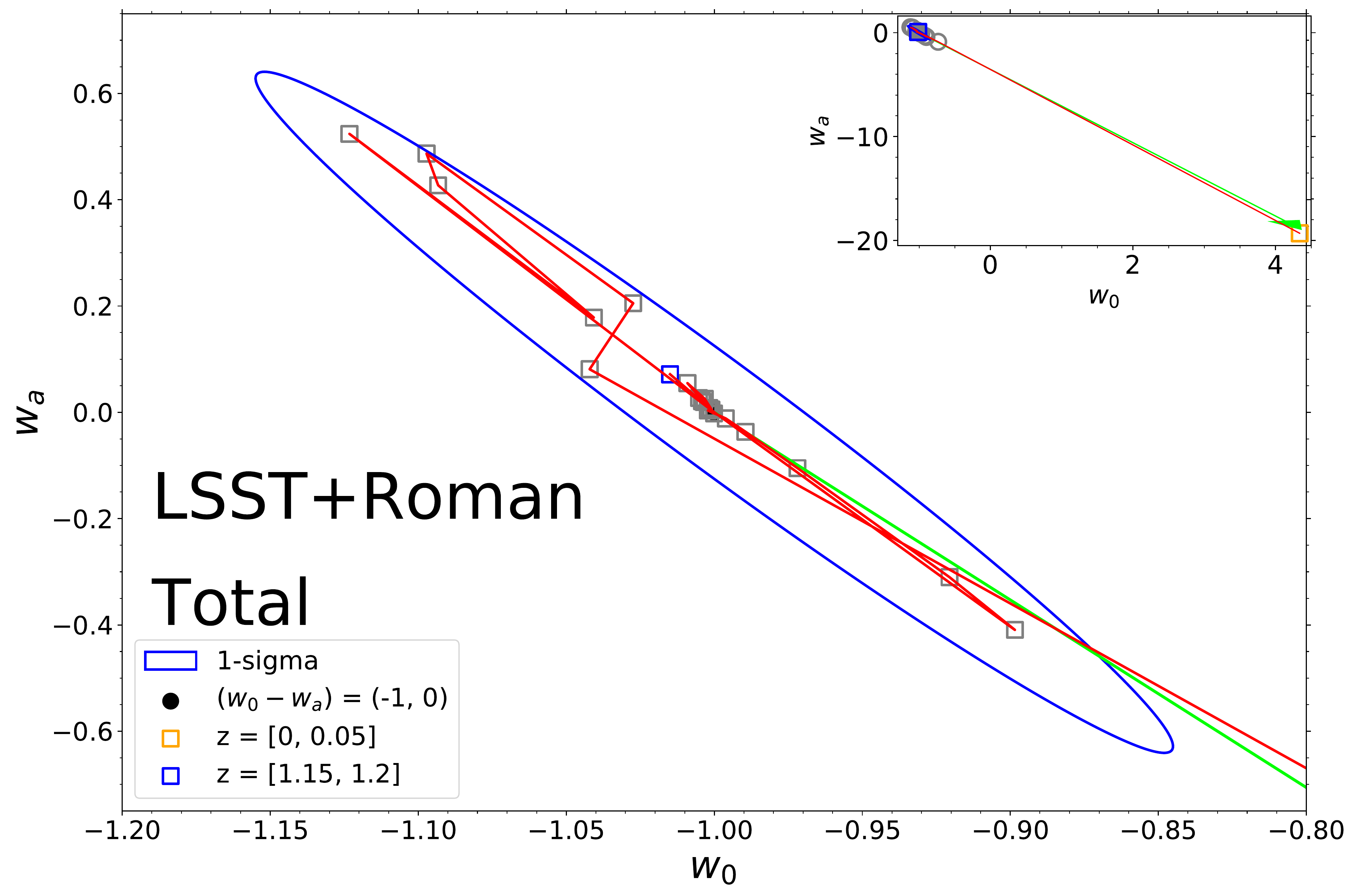}%
    \caption{As Figure~\ref{Vd}, but for LSST+Roman. 
}
    \label{Ve}
\end{figure}

Figures~\ref{Vd} and \ref{Ve} show the total cosmology bias 
from all galaxy photo-z systematics for LSST alone and for 
LSST+Roman. Again, qualitatively they are similar and 
quantitatively there is not a large difference. 
While cosmology bias is significantly reduced around $z\approx0.5$ 
(by at least $\delta w_a>0.15$), due to a combined 
reduction in photo-z outlier and inlier systematics there, and 
photo-z systematics is improved at $z\gtrsim1$ -- but systematics 
there causes relatively little cosmology bias -- the low redshift 
systematics remains, and this substantially drives the cosmology  
bias. 

Thus, the addition of NIR data does not obviate the need for 
spectroscopic redshifts for SN host galaxies at $z\lesssim0.2$, 
and further improvements around $z\approx0.5$ would be useful 
as well.

\section{Conclusions}\label{c4} 

Supernovae continue to be one of the most incisive probes 
of cosmic acceleration. With forthcoming surveys the data 
available will vastly increase, limited however by spectroscopic 
followup. We consider the use of Type Ia supernovae with only 
photometric redshifts, assessing the impact of systematics in 
the redshift estimation through large simulations of host galaxy  
colors. This builds on the analytic work of \cite{f1} by 
using mock data meant to emulate LSST observations to derive 
the systematics requirements for controlling bias in the 
cosmological parameter estimation, particularly for dark 
energy. 

Simulating half a million galaxies provides us with good 
statistics on the redshift systematics, categorized into 
outlier and core systematics, as a function of true redshift. 
We propagate this to cosmological parameter bias for a 
LSST-like survey, showing the impact of each individual 
redshift range as well as the full sum. Our results show 
that for both outliers and inliers, the redshift systematic requirement is  
reduction by an order of magnitude -- principally in the lower 
redshift range -- for the bias not to exceed the 68\% 
confidence statistical uncertainty in the $w_0$--$w_a$ plane. 

The low redshift photometric systematics are the most dangerous. 
Fortunately, they are also easiest to mitigate with further observations. The favored 
situation would be to use the photometric supernova for cosmology only for $z>0.3$, and obtain  
spectroscopic followup for all supernovae at $z\lesssim0.2-0.3$ (see below). 

We explore three potential methods for controlling 
systematics: modeling, select spectroscopy, and external 
imaging data. Fully successful modeling, i.e.\ a residual 
at only the level of 
the bootstrap uncertainty on a large suite of simulations, 
would be ideal, while a residual an order of 
magnitude below base is 
necessary. 
Improvement in understanding host galaxy properties 
(for both the core bias and the outlier bias) is desirable, 
e.g.\ are the spectroscopic catalogs used 
for training the CMNN estimators -- and the 
resulting outliers -- representative, 
in particular of Type Ia SN host galaxies. 
Machine learning algorithms to estimate the redshifts will 
likely continue to become better; Ref.~\cite{cmnn2} has 
summarized the accuracy for several different algorithms, 
and while the CMNN estimator fares quite well in comparison 
to most of the other machine learning estimators, the authors of \cite{m2} state that 
it is not optimized for the absolute best fit, but rather 
aims to assess differences in survey strategy. 

Select spectroscopy is an attractive and highly practical 
solution. This would involve a multiobject spectrograph 
survey to obtain SN host galaxy redshifts out to $z\lesssim0.2$, 
where the systematics have the greatest impact on cosmology 
bias. 
Such data has numerous other astrophysical applications, 
including for probing gravity with peculiar velocities 
\cite{pecvel,peculiar}. 

External imaging data will be available from NIR surveys such as 
the Nancy Grace Roman Space Telescope and the Euclid 
satellite. We carry out an analysis including Roman NIR 
filter constraints on galaxy photometric redshifts with 
LSST mock data in the simulations. We consider the impact 
on systematics from photo-z outliers, inliers (the 
complement of outliers), the total set, and a special 
like to like comparison where we choose the same set of 
galaxies from both LSST and LSST+Roman. Especially for 
the like to like comparison, Roman helps in controlling the 
systematics. However, overall the main redshift estimation 
improvement is at high redshifts where the cosmology bias 
is less. The sensitive low redshifts show relatively little 
gain. So even in the era of LSST, Euclid, and Roman, low redshift 
spectroscopy of SN host galaxies will be quite important for 
enabling the full potential of dark energy  constraints from 
supernova cosmology through time domain surveys.

\acknowledgments 
We are very grateful to Melissa Graham for making her data and code public, 
and many useful tips on how to adapt them. We also thank 
Rick Kessler, Alex Kim, Gautham Narayan, and the LSST DESC Supernova 
and Photometric Redshift working groups for comments and 
suggestions. AM acknowledges the support of ORAU Grant No.~110119FD4534.
EL is supported in part by the 
U.S.\ Department of Energy, Office of Science, Office of High Energy 
Physics, under contract no.~DE-AC02-05CH11231, by NASA ROSES grant 12-EUCLID12-0004, and by 
the Energetic Cosmos Laboratory.

\bibliographystyle{unsrt}
\bibliography{main}

\end{document}